\documentclass[%
 reprint,
superscriptaddress,
nofootinbib,
 amsmath,amssymb,
 aps,
floatfix,
]{revtex4-2}
\usepackage{booktabs}
\usepackage{graphicx}
\usepackage{bm}
\usepackage{amsmath,amsfonts,amssymb,slashed,upgreek,color}
\usepackage{xcolor}
\usepackage{hyperref}
\usepackage{cancel}
\graphicspath{{plots/}}
\usepackage{siunitx}
\usepackage[flushleft]{threeparttablex}

\usepackage{blindtext}
\usepackage{float}
\usepackage{multirow}
\newcommand{\Cgh}{C_\ell^{{\rm g} \rm{HI}}}
\newcommand\ee{\end{equation}}
\newcommand\be{\begin{equation}}

\newcommand{\bn}{\mathbf{n}}
\newcommand{\deltaim}{\Delta_{\rm HI}}
\newcommand{\ec}{E^{\times}_\ell}

\newcommand{\hec}{\hat{E}_\ell^{\times}}

\newcommand{\im}{{\rm HI}}

\newcommand{\g}{{\rm g}}
\newcommand{\hEG}{\hat{E}_G}

\hypersetup{pdfpagemode=UseNone, pdfstartview=FitH, linkcolor=linkblue, 
            citecolor=linkgreen, urlcolor=linkblue}

\newcolumntype{C}[1]{>{\centering\arraybackslash}p{#1}}
\newcommand\Tstrut{\rule{0pt}{2.6ex}}       % Top strut
\begin{document}

%\preprint{APS/123-QED}

\title{Model-Independent Test for Gravity using Intensity Mapping and Galaxy Clustering} 

\author{Muntazir M. Abidi}
\affiliation{%
Universit\'e de Gen\`eve, D\'epartement de Physique Th\'eorique and Centre for Astroparticle Physics, 24 quai Ernest-Ansermet, CH-1211 Gen\`eve 4, Switzerland
}%
\author{Camille Bonvin}
\affiliation{%
Universit\'e de Gen\`eve, D\'epartement de Physique Th\'eorique and Centre for Astroparticle Physics, 24 quai Ernest-Ansermet, CH-1211 Gen\`eve 4, Switzerland
}%
\author{Mona Jalilvand}
\affiliation{%
Universit\'e de Gen\`eve, D\'epartement de Physique Th\'eorique and Centre for Astroparticle Physics, 24 quai Ernest-Ansermet, CH-1211 Gen\`eve 4, Switzerland
}%
\affiliation{%
Department of Physics, McGill University, 3600 rue University, Montreal, QC H3A 2T8, Canada
}%
\affiliation{%
McGill Space Institute, McGill University, 3550 rue University, Montreal, QC H3A 2A7, Canada
}%

\author{Martin Kunz}
\affiliation{%
Universit\'e de Gen\`eve, D\'epartement de Physique Th\'eorique and Centre for Astroparticle Physics, 24 quai Ernest-Ansermet, CH-1211 Gen\`eve 4, Switzerland
}%

\date{\today}% 

\begin{abstract}
We propose a novel method to measure the $E_G$ statistic from clustering alone. The $E_G$ statistic provides an elegant way of testing the consistency of General Relativity by comparing the geometry of the Universe, probed through gravitational lensing, with the motion of galaxies in that geometry. Current $E_G$ estimators combine galaxy clustering with gravitational lensing, measured either from cosmic shear or from CMB lensing. In this paper, we construct a novel estimator for $E_G$, using only clustering information obtained from two tracers of the large-scale structure: intensity mapping and galaxy clustering. In this estimator, both the velocity of galaxies and gravitational lensing are measured through their impact on clustering. We show that with this estimator, we can suppress the contaminations that affect other $E_G$ estimators and consequently test the validity of General Relativity robustly. We forecast that with the coming generation of surveys like HIRAX and \emph{Euclid}, we will measure $E_G$ with a precision of up to 7\% (3.9\% for the more futuristic SKA2).

\end{abstract}

\pacs{Valid PACS appear here}% PACS, the Physics and Astronomy
                             
\maketitle

%\tableofcontents

%=========================

\section{Introduction}

One of the main goals of the coming generation of large-scale structure surveys is to test the consistency of General Relativity (GR) at cosmological scales. Since the observation of the accelerated expansion of the Universe in 1998 \citep{Riess:1998fmf, Perlmutter:1998vns}, a large number of theories of gravity beyond General Relativity have been constructed. Testing these theories one by one by confronting them with observations is not anymore a feasible option. This complexity in the landscape of models beyond GR has led the community to build model-independent tests of gravity, i.e.\ tests that can be applied to data without relying on particular modelling and whose outcome will either confirm or rule out the validity of GR (e.g.\ \citep{Kunz:2012aw,Amendola:2012ky,Pinho:2018unz,Franco:2019wbj,Bonvin:2020cxp,Boschetti:2020fxr,Sobral-Blanco:2021cks,Raveri:2021dbu}).

One particularly useful test is the so-called $E_G$ statistic, first proposed by Zhang et al.~\citep{Zhang:2007nk}. The idea of this test is to combine galaxy-lensing correlations with galaxy-velocity correlations to test the relation between the sum of the metric potentials, $\Phi+\Psi$, and the galaxy peculiar velocity $V$:    
\be
E_G=\frac{\rm{galaxy-lensing}}{\rm{galaxy-velocity}}\propto \frac{\langle \delta_\g (\Phi+\Psi)\rangle}{\langle \delta_\g V\rangle}\, . \label{eq:EGdef}
\ee
In GR, these two quantities are related via Einstein's equations so that $E_G$ takes a specific scale-independent value. In modified theories of gravity, however, the relation between $\Phi$ and $\Psi$ is generically modified, as is the growth rate of structures $f$, which governs the evolution of peculiar velocities \citep{Pogosian2008PhRvD}: $E_G$ is consequently modified and becomes potentially scale-dependent. Measurements of $E_G$ provide, therefore, a direct test of the validity of GR. This test has the advantage of being independent of galaxy bias $b$, since the galaxy density appears both in the numerator and denominator of Eq.~\eqref{eq:EGdef}, and also of the initial conditions, which cancel out in the ratio \cite{Amendola:2012ky}.

Different methods have been used to measure $E_G$ in practice. First, since peculiar velocities are not straightforward to measure, the galaxy-velocity correlation in the denominator has been replaced by the product of the galaxy-galaxy correlations and the parameter $\beta=f/b$, which can be measured directly from redshift-space distortions (RSD)~\citep{Kaiser1987}. Concerning the numerator, two observables have been used: 1) the correlations between galaxy clustering and cosmic shear, and 2) the correlations between galaxy clustering and CMB lensing.

The first measurement of $E_{G}$ was carried out by Reyes et al.~\citep{Reyes:2010tr} using the first method applied to luminous red galaxies from the Sloan Digital Sky Survey (SDSS)~\citep{York2000}. They measured $E_G = 0.39 \pm 0.06$, confirming the $\Lambda$CDM predictions, on the scales of tens of Mpc. Later, Amon et al.~\citep{Amon2018MNRAS.479.3422A} measured $E_G$, again with the first method, by combining deep imaging data from the Kilo-Degree Survey \citep{Kuijken:2015vca} with the overlapping spectroscopic 2-degree Field Lensing Survey~\citep{Blake2016MNRAS.462.4240B}, the Baryon Oscillation Spectroscopic Survey~\citep{Dawson2013AJ} and the Galaxy Mass Assembly Survey~\citep{Driver2011MNRAS} and found some tension of their results with the GR predictions. Pullen et al.~\citep{Pullen:2015vtb} used the second method, cross-correlating the Planck CMB lensing map with the SDSS III CMASS galaxy samples and found an almost 2.6$\sigma$ deviation from the GR prediction. Other works on the $E_G$ measurements found similar tensions~\citep{Alam:2016qcl, Blake2015, delaTorre:2013rpa}.

Measuring $E_G$ more precisely with the coming generation of surveys will reveal whether these tensions persist, possibly indicating a breakdown of GR at cosmological scales. In this context, minimising the impact of systematic effects on measurements of $E_G$ is crucial. Recently, it has been shown that one important contamination, which is negligible for current surveys, will contaminate $E_G$ at high redshift, invalidating its use to test the consistency of GR with the next generation of surveys~\cite{MoradinezhadDizgah:2016pqy}. This contamination is the contribution of lensing magnification to galaxy clustering. Since lensing magnification correlates strongly with cosmic shear (method 1) and with CMB lensing (method 2), it inevitably leads to an extra contribution in the numerator of Eq.~\eqref{eq:EGdef}. At $z=1.5$ this contamination reaches 25-40\%, leading to an $E_G$ which is neither scale-independent nor bias-independent. In~\cite{Ghosh2019a} a method has been proposed to remove this contamination by measuring additional correlations, namely shear-shear correlations and shear-CMB lensing correlations, and subtracting them from $E_G$.

In this paper, we propose an alternative way of measuring $E_G$ \emph{without contamination}, using only clustering information. We use two different tracers of the large-scale structure (LSS): galaxy clustering and 21\,cm intensity mapping (IM). Intensity mapping is a novel technique to map the LSS by measuring the intensity fluctuations of some emission line (typically the 21\,cm line emitted by neutral hydrogen) with radio surveys. These fluctuations, which follow the dark matter distribution, can be used as a new tracer of the LSS. Although the presence of large foregrounds make IM auto-correlation measurements very challenging, this problem is mitigated in cross-correlation measurements of IM with galaxy clustering, which have already been successfully performed ~\cite{Chang:2010jp,Masui:2012zc,Switzer:2013ewa,Anderson:2017ert,Tramonte:2020csa,Li:2020pre,Wolz:2021ofa,CHIME:2022kvg,Cunnington:2022uzo}.

In~\cite{Jalilvand2019}, we built a new observable, called GIMCO, which combines IM with galaxy clustering to obtain a direct measurement of the galaxy-lensing correlation. Here we propose to use GIMCO in the numerator of $E_G$. Instead of measuring the galaxy-lensing correlation from cosmic shear or CMB lensing, we measure it directly from clustering. As we will see, this method has the strong advantage of being unaffected by the lensing contamination described above, which affects both galaxy-shear correlations and galaxy-CMB lensing correlations. Moreover, since it relies only on clustering information, it is unaffected by potential inconsistencies between low redshift and high redshift data sets (as is the case for galaxy-CMB lensing correlations) or systematics affecting the measurement of cosmic shear. It provides, therefore, a robust way of testing the consistency of GR.

The rest of the paper is organised as follows: In Sec.~\ref{EG_theory} we introduce our new method to measure $E_G$ using intensity mapping. In Sec.~\ref{sec:SNR} we study the contaminations to $E_G$, and we determine the optimal redshift binning to reduce these contaminations to a negligible level. In Sec.~\ref{sec:forecasts} we forecast the precision with which $E_G$ will be measured with the coming generation of galaxy and IM surveys and the constraints expected on modifications of gravity. We conclude in Sec.~\ref{sec:conclusion}.

%=========================
\section{ ${\emph E}_G$ from intensity mapping}
\label{EG_theory}

We build $E_G$ using two different biased tracers of the matter clustering: the fluctuations in galaxy number counts, $\Delta_{\rm g}$, and the fluctuations in the 21\,cm brightness temperature, $\deltaim$. The fully relativistic expressions at linear order for these two quantities have been derived in~\cite{Yoo:2009au,Bonvin:2011bg,Challinor:2011bk,Hall:2012wd}. Here we consider only the terms that dominate in the angular power spectrum of thick redshift bins. For the galaxy number counts, we have 

\begin{equation}
    \Delta_{\rm g}(\mathbf{n}, z) = b_{\rm g}(z)\delta(\mathbf{n},z) + \big(5s(z)-2\big)\kappa(\mathbf{n},z)\, ,
    \label{eq:numbercount}
\end{equation}
where $\bn$ denotes the direction of observation, $z$ is the redshift, $b_{\rm g}$ is the linear galaxy bias, $\delta$ is the matter density contrast, and $s$ is the magnification bias. The first term in Eq.~\eqref{eq:numbercount} is the density contribution, and the second term is the lensing magnification, proportional to the lensing convergence $\kappa$,
\be
\kappa(\mathbf{n}, z)= \frac{1}{2}\int_{0}^{\chi(z)}\! d\chi' \,\frac{\chi(z) -\chi' }{\chi(z) \chi'} \, \nabla^2_{\Omega} (\Phi+\Psi)(\mathbf{n},\chi')\, ,
\label{eq:kappa}
\ee
where $\nabla^2_{\Omega}$ denotes the angular Laplacian~\footnote{Note that we use the following metric convention $\text{d}s^2 = a(\eta)^2\big[-(1+2\Psi)\text{d}\eta^2 + (1-2\Phi)\delta_{ij}\text{d}x^i\text{d}x^j\big]$, where $\eta$ denotes conformal time, so that $\chi=\eta_0-\eta$ for the conformal time today $\eta_0$.}. 

Contrary to galaxy number counts, the fluctuations in the brightness temperature are not affected by lensing magnification at linear order in perturbation theory due to conservation of surface brightness~\cite{Hall:2012wd}, and we have
\be
\deltaim(\mathbf{n}, z) = b_{\rm HI}(z)\delta(\mathbf{n},z)\, , \label{eq:IMcount}
\ee
where $b_\im$ is the bias of neutral hydrogen. In Eqs.~\eqref{eq:numbercount} and~\eqref{eq:IMcount} we have neglected redshift-space distortions (RSD) since they are subdominant for thick redshift bins. In our forecasts, we will choose the binning such that this contribution is always negligible and does not contaminate $E_G$, see Section~\ref{sec:SNR}.

We expand the galaxy number counts and brightness temperature in spherical harmonics 
\begin{align}
&  \Delta_{\rm g}(\bn,z) = \sum_{\ell m}a_{\ell m}^{\rm g}(z)Y_{\ell m}(\boldsymbol{n})\, ,\\
& \deltaim(\bn,z) = \sum_{\ell m}a_{\ell m}^{\rm HI}(z)Y_{\ell m}(\boldsymbol{n})\, ,
\end{align}
and consider their angular power spectra
\begin{equation}
    C^{\text{XY}}_{\ell}(z,z') = \big\langle a_{\ell m}^{\text{X}}(z)a_{\ell m}^{*\text{Y}}(z')\big\rangle\, ,
\end{equation}
for $\rm{X,Y}=g, {\im}$. We can then build $E_G$ from these spectra in the following way: the numerator of $E_G$ is given by the galaxy-lensing correlation. Since galaxy number counts are affected by lensing magnification whereas intensity mapping is not, the following estimator, proposed for the first time in~\cite{Jalilvand2019}, is directly proportional to the galaxy-lensing correlation
\begin{align}
E^\times_{\ell}(z_f, z_b) &\equiv C_\ell^{\im {\rm g}}(z_f, z_b)-C_\ell^{{\rm g}\im}(z_f, z_b)\label{eq:El}\\
&= \big(5 s(z_b)-2\big)b_{\im}(z_f)C_{\ell}^{\delta\kappa}(z_f, z_b) \nonumber\\
&-\big(5s(z_f)-2\big)b_{\im}(z_b)C_{\ell}^{\kappa\delta}(z_f, z_b)\nonumber\\
&+\big[b_{\text{g}}(z_b)b_{\im}(z_f)-b_{\text{g}}(z_f)b_{\im}(z_b)\big] C_{\ell}^{\delta \delta}(z_f, z_b)\, . \nonumber
\end{align}
The term we are interested in is the one in the second line, $C_{\ell}^{\delta\kappa}(z_f,z_b)$, which represents the lensing magnification of background galaxies generated by a foreground density at $z_f$. The other lensing term in the third line, $C_{\ell}^{\kappa\delta}(z_f, z_b)$, is always negligible since it is due to correlations of foreground lensing magnification with background density. Finally, the density term in the last line is suppressed by two effects: first by the bias difference, which would exactly vanish if the biases were redshift independent, and second by the fact that density correlations, $C_{\ell}^{\delta \delta}(z_f, z_b)$, quickly decrease with redshift separation. In Sec.~\ref{sec:contaminations} we will choose the redshift bins such that this density contamination is negligible. This estimator, called GIMCO~\cite{Jalilvand2019}, provides therefore a robust way of isolating lensing magnification, and we will use it in the numerator of $E_G$
\be
E^\times_{\ell}(z_f, z_b) \simeq   \big(5 s(z_b)-2\textbf{})b_{\im}(z_f)C_{\ell}^{\delta\kappa}(z_f, z_b)\, .\label{eq:GIMCO_app}
\ee

In the denominator of $E_G$ we need the galaxy-velocity correlation. In~\cite{Pullen2016}, this correlation was replaced by the product of $\beta_{\rm g}(z_f)=f(z_f)/b_{\rm g}(z_f)$ and the galaxy-galaxy correlations. In our case, instead of the galaxy-galaxy correlations, we use the galaxy-intensity mapping correlations, such that the $\im$ bias in Eq.~\eqref{eq:GIMCO_app} cancels. We therefore have
\begin{align} 
E_{G}&=\frac{E_\ell^\times(z_f,z_b)}{\beta_{\g}(z_{f}) \Cgh \left(z_{f}, z_{f}\right)}= \frac{ C_\ell^{{\rm HI} \rm g} (z_{f}, z_{b}) - \Cgh (z_f,z_b)}{\beta_{\rm g}\left(z_{f}\right) \Cgh\left(z_{f}, z_{f}\right)}\nonumber\\
&=\frac{\big(5 s(z_b)-2\big)C_{\ell}^{\delta\kappa}(z_f, z_b)}{f(z_f)C_{\ell}^{\delta\delta}(z_f, z_f)}\, .
\label{eq:Eg}
\end{align}
We see from Eq.~\eqref{eq:Eg} that we only need two types of correlators to measure $E_G$: the cross-correlation between galaxy clustering and intensity mapping and the auto-correlation of galaxy clustering, from which $\beta_\g$ is measured. This is particularly interesting to test deviations from GR: Eq.~\eqref{eq:Eg} tests the relation between density, velocity and lensing potential measured from the same correlators. Therefore, the outcome of this test is not subject to potential inconsistencies between different data sets, as can be the case for standard versions of $E_G$ that rely on galaxy clustering and shear or on galaxy clustering and CMB lensing. 

As discussed above, the numerator in Eq.~\eqref{eq:Eg} is not affected by the lensing contamination computed in~\cite{MoradinezhadDizgah:2016pqy}: the density-magnification contribution is the \emph{signal} in our estimator, whereas the magnification-magnification contamination is absent since intensity mapping is insensitive to lensing magnification. The denominator in Eq.~\eqref{eq:Eg} contains density-magnification contamination and an RSD contamination. In Section \ref{sec:SNR} we will show that we can choose the redshift binning such that these two contaminations remain negligible. 

Note that an $E_G$ estimator using intensity mapping has already been proposed in~\cite{Pourtsidou_2016}. However, this estimator differs from ours: in~\cite{Pourtsidou_2016} the lensing signal is not measured from clustering as we do here, but from CMB lensing cross-correlated with intensity mapping. This estimator, like ours, is unaffected by magnification contamination. It relies, however on the auto-correlation of intensity mapping to measure $\beta_{\im}$ as well as $C_\ell^{\im\, \im}$. The auto-correlations of intensity mapping will be very challenging to measure due to the difficulty of subtracting the foregrounds accurately. Our estimator relies instead only on cross-correlations of intensity mapping with galaxy clustering, which are unaffected by intensity mapping foregrounds and have already been measured~\cite{Chang:2010jp,Masui:2012zc,Switzer:2013ewa,Anderson:2017ert,Tramonte:2020csa,Li:2020pre,Wolz:2021ofa,CHIME:2022kvg,Cunnington:2022uzo}.

\section{Signal-to-Noise and Contamination}
\label{sec:SNR}

We now study the detectability of $E_G$ with the coming generation of intensity mapping and galaxy surveys. For intensity mapping, we consider the 21\,cm intensity mapping survey HIRAX, which will measure the neutral hydrogen distribution in the redshift range of $z=0.775$ to $2.55$ covering 15'000 square degrees of the southern sky~\citep{Newburgh:2016mwi,Crichton:2021hlc}. For the galaxy survey, we study two examples, one modelled on SKA phase 2, covering 30'000 square degrees, based on the specifications in ~\cite{Villa:2017yfg, Bull:2015lja}, and a 15'000 square-degree \emph{Euclid}-like survey~\cite{Euclid:2021icp}, based on~\cite{Euclid:2019clj,Euclid:2021rez}.
We perform separate forecasts for the spectroscopic and photometric samples for the \emph{Euclid}-like scenario. The spectroscopic sample has the advantage of providing an accurate measurement of $\beta_\g$, through RSD, whereas the photometric sample will measure $\beta_\g$ with relatively large error bars. On the other hand, the photometric sample has a significantly lower shot noise due to the much larger number of galaxies detected and extends to higher redshifts. It is, therefore, interesting to study how the measurement of $E_G$ differs in these two samples. The overlapping redshift range between HIRAX and the galaxy surveys are $z\in [0.9, 1.8]$ for \emph{Euclid}-like spectroscopic, $z\in [0.78, 2.16]$ for \emph{Euclid}-like photometric and $z\in [0.78, 2.0]$ for SKA2-like. We assume that \emph{Euclid}-like has a $2/3$ aerial overlap with HIRAX, leading to $f_{\text{sky}}=0.242$, whereas the SKA2-like survey fully overlaps with HIRAX leading to $f_{\text{sky}}=0.363$.

For HIRAX correlated with the \emph{Euclid}-like spectroscopic and the SKA2-like surveys, we consider top-hat redshift bins since the accuracy in the redshift determination is excellent. For the \emph{Euclid}-like photometric surveys, on the other hand, we use the bins given in~\cite{Euclid:2021rez} (see Fig.\ 3), which we approximate with Gaussian window functions according to Table 4 in~\cite{Nistane:2022xuz}. Since we are using the code {\sc class}~\citep{class2,DiDio:2013bqa} to compute the angular power spectra, and it does not allow us to choose two different windows at two different redshifts, we also use Gaussian windows for HIRAX in our forecasts in this case.

The signal depends on the galaxy magnification bias, $s$. For \emph{Euclid}-like photometric, we use the magnification bias measured from the flagship simulation, see Table 1 of~\citep{Euclid:2021rez}. For SKA2, we use the expression from \citep{Jelic-Cizmek:2020pkh}, and for \emph{Euclid}-like spectroscopic, we use the model developed in~\cite{Montanari:2015rga}. The signal is by construction independent of galaxy bias, however, the variance of $E_G$ depends on the galaxy bias. Explicit expressions for the biases used to model the different surveys can be found in Appendix~\ref{app:surveys}.

\subsection{Limber approximation}

$E_G$ depends on the foreground and background redshifts, $z_f$ and $z_b$. However, in Limber approximation, the background dependence goes away, and $E_G$ directly probes deviations from GR at redshift $z_f$. We first relate the metric potentials (that enter in the convergence) to the matter density contrast, $\delta$, using the Poisson equation. Since we want to test the validity of GR, we allow for deviations, encoded in the parameter $\Sigma$, see e.g.~\cite{Kunz:2006ca,Amendola:2007rr,Pogosian:2010tj,Kunz:2012aw}
\begin{equation}
k^2(\Phi + \Psi ) =- 3 H^2_0\Omega_{\text{m}}(1+z)\Sigma(z)\delta(k,z)\, .
\label{eq:mod_Poisson}
\end{equation}
Here $\Omega_{\text{m}}$ denotes the matter density parameter today, and $H_0$ represents the present-time value of the Hubble parameter.
In $\Lambda$CDM, $\Sigma=1$ on the scales and redshifts of interest for us. Deviations from $\Lambda$CDM can be encoded into a modification to the Poisson equation, and a difference between the two metric potentials, which combine into a $\Sigma$ generically different from 1~\cite{Kunz:2006ca,Amendola:2007rr,Pogosian:2010tj,Kunz:2012aw}. Plugging this into the convergence~\eqref{eq:kappa}, and using Limber approximation~\cite{Limber:1954zz} we obtain the density-magnification correlation, see e.g.~\cite{Euclid:2021rez}
\begin{align}
    &C_{\ell}^{\delta \kappa}(z_f,z_b) = \frac{\ell(\ell+1)}{(\ell+1/2)^2}\frac{3H_0^2 \Omega_{\text{m}}}{2}\int_{z^{\rm min}_f}^{z^{\rm max}_f}\!\!\!\text{d}z\,n_{\im}(z) \Sigma(z)\nonumber\\
    &\times(1+z)P_{\delta\delta}\left[\frac{\ell+1/2}{\chi(z)},z\right]\int_{z^{\rm min}_b}^{z^{\rm max}_b}\!\!\!\text{d}z'\,n_{\g}(z') 
    \frac{\chi(z')-\chi(z)}{\chi(z)\chi(z')} 
    \label{eq:clkg}
\end{align}
where $n_\g(z)$ is the normalised galaxy distribution, $n_\im(z)$ is the normalised intensity distribution and the integrals run over the size of the redshift bins. Note that the $\ell$-dependent coefficient in front of the integral can be set to one at large $\ell$, where Limber approximation is valid. Moreover, we assume that $\Sigma$ is independent of $k$, which is a common assumption in large-scale structure analyses (e.g.~\cite{Planck:2015bue}), and is motivated by the fact that various theories of modified gravity obey this assumption in the quasi-static approximation, e.g.~\cite{Brans:1961sx,Dvali:2000hr,Babichev:2009ee}. 

The density-density correlation can also be simplified using Limber approximation, leading to
\begin{equation}
    C^{\delta\delta}_{\ell}(z_f,z_f) = \int_{z^{\rm min}_f}^{z^{\rm max}_f}\!\!\!\text{d}z\, \frac{H(z)}{\chi^2(z)}n_{\g}(z)n_\im(z)P_{\delta\delta}\left[\frac{\ell+1/2}{\chi(z)},z\right]
     \label{eq:cldd}
\end{equation}
Inserting Eqs.~\eqref{eq:clkg} and~\eqref{eq:cldd} in $E_G$~\eqref{eq:Eg} and assuming that the functions $\Sigma(z), \chi(z), H(z)$ and $n_\g(z)$ vary slowly with redshift inside each redshift bin, we obtain
\begin{align}
    E_G(\ell, z_f, z_b) = \Gamma(z_f, z_b)\frac{\Omega_m \Sigma(z_f)}{f(z_f)}\, ,
\end{align}
where
\begin{align}
\Gamma(z_f,z_b)\equiv& \big(5s(z_b)-2\big)\frac{3 H_0^2}{2H(z_f) }(1+z_f) \\ 
& \times \int_{z^{\rm min}_f}^{z^{\rm max}_f}\!\!\!\text{d}z\,n_{\im}(z)
    \int_{z^{\rm min}_b}^{z^{\rm max}_b}\!\!\!\text{d}z'\,n_{\g}(z') 
    \frac{\chi(z')-\chi(z)}{\chi(z)\chi(z')} \nonumber \\
& \times \left( \int_{z^{\rm min}_f}^{z^{\rm max}_f}\!\!\!\text{d}z\,n_{\g}(z)n_\im(z) \right)^{-1}\nonumber \\
\simeq&\big(5s(z_b)-2\big)\frac{3 H_0^2}{2H(z_f) }(1+z_f)\label{eq:gamma}\\
&\times\frac{n_\g(z_b)}{n_\g(z_f)}\frac{1}{\chi(z_f)}\int_{z^{\rm min}_b}^{z^{\rm max}_b}\!\!\!\text{d}z\frac{\chi(z)-\chi(z_f)}{\chi(z)}\, .\nonumber
\end{align}
The second expression corresponds to top-hat bins, while the first one allows for more general redshift bins.

The coefficient $\Gamma$ depends on the background cosmology, the galaxy distribution and the magnification bias. A common assumption in large-scale structure analyses is to fix the background to $\Lambda$CDM (since the CMB has extremely well constrained it) and to test for deviations from General Relativity at the level of the perturbations. This is, for example, the strategy adopted in RSD analyses to measure $\beta_\g$ in a model-independent way, see e.g.~\cite{BOSS:2016psr}. Here we follow the same procedure. The galaxy distribution and the magnification bias are quantities that can directly be measured for a given population of galaxies and a given survey. Consequently, $\Gamma$ is a parameter we can predict. Note that we do not include $\Omega_{\text{m}}$ in $\Gamma$ as it cannot be measured in a model-independent way when allowing for general dark energy or modified-gravity scenarios \cite{Kunz:2007rk,Amendola:2012ky}. We then define a re-scaled $E_G$ variable, by dividing it with $\Gamma (z_f, z_b)$
\begin{align}
\hat{E}_G(\ell, z_f, z_b)\equiv&\frac{E_\ell^\times(z_f,z_b)}{\Gamma(z_f,z_b)\beta_\g(z_f)C_\ell^{\g\im}(z_f,z_f)}\nonumber \\
=&\frac{\Omega_{{\text{m}}}\Sigma(z_f)}{f(z_f)}\, .   \label{eq:EGhat}
\end{align}

\begin{figure}
    \centering
    \includegraphics[width = \linewidth]{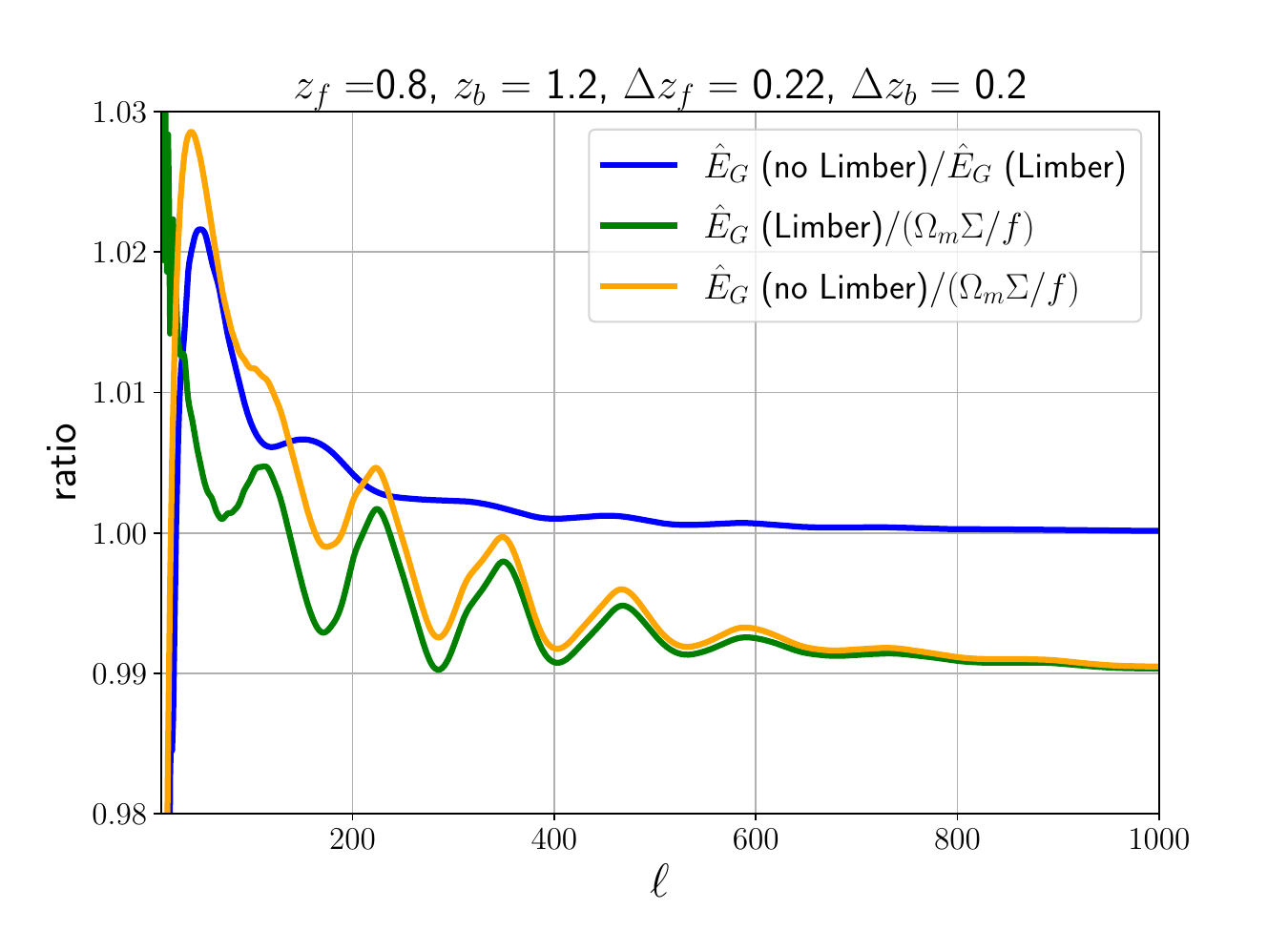}
    \caption{Accuracy of the approximations used in our estimator: in blue, we show the ratio of $\hEG$ computed without and with Limber approximation; in green the ratio with and without redshift evolution; and in orange the ratio without and with both approximations.}
    \label{fig:EGratioPlots}
\end{figure}

We have tested the validity of the approximations used to obtain Eq.~\eqref{eq:EGhat}. In Fig.~\ref{fig:EGratioPlots} we compare $\hEG$ using the full-sky angular power spectra, with the result using Limber approximation, and with the second line of Eq.~\eqref{eq:EGhat}, which further neglects redshift evolution within each bin. Limber approximation induces a sub percent error on $\hEG$ for $\ell$ above 100 (and less than $0.1\%$ for $\ell \geq 400$). Neglecting redshift evolution induces a $\sim1\%$ percent error at all $\ell$'s. This is well below the statistical uncertainties on $\hEG$, as we will see in Section~\ref{sec:forecasts}. The model-independent and $\ell$-independent function $\Gamma(z_f,z_b)$ defined in Eq.~\eqref{eq:gamma} is therefore perfectly adapted and allows us to define $\hEG$ in a model-independent way, in contrast to the function $C_\Gamma(\ell)$ used in~\cite{Pullen:2015vtb}, which depends on the density power spectrum and is thus model-dependent.~\footnote{Note that to properly test the validity of Eq.~\eqref{eq:EGhat}, it is necessary to use the same distribution functions, $n_\g$ and $n_\im$, in Eqs.~\eqref{eq:clkg}, \eqref{eq:cldd} and~\eqref{eq:gamma}. In our case, we use top-hat window functions approximated by a smooth function in CLASS, with a sharpness that can be adjusted.}

\subsection{Signal-to-noise ratio}
We now compute the signal-to-noise ratio (SNR) of $\hEG$ for the three combinations of surveys. Since the signal is independent of the background redshift, as discussed above, we sum over all pairs at fixed $z_f$:

\begin{align}
\label{eq:sn_signal}
&\left(\frac{\rm S}{\rm N}\right)^2(z_f)=\sum_{\ell,\ell'=\ell_{\rm min}}^{\ell_{\rm max}}\sum_{z_b, z_b'=z_{\rm min}}^{z_{\rm max}}\hat{E}_G(\ell, z_f, z_b) \\
&\times {\rm cov}^{-1} \big[\hat{E}_G(\ell, z_f, z_b),\hat{E}_G(\ell', z_f, z_b')\big]\hat{E}_G(\ell', z_f, z_b')\, .\nonumber
\end{align}
The sum runs over all multipoles used in the analysis and all background redshift bins. In Section~\ref{sec:contaminations} we will select those such that the contamination from RSD, lensing and density are negligible.

The full expression for the covariance of $\hEG$ is given in Appendix~\ref{app:covariance}. It depends on the covariance of $C_\ell^{\g\im}$, on the variance of $\beta_\g$ and the covariance between the two. The covariance of $C_\ell^{\g\im}$ is given by
\begin{align}
&{\rm cov}\left[C_\ell^{\g\im}(z_1,z_2), C_\ell^{\g\im}(z_3,z_4)\right]=\frac{1}{f_{\rm sky}(2\ell+1)} \label{eq:covClgim}\\
&\times\left[C_\ell^{\g\g}(z_1,z_3)C_\ell^{\im\im}(z_2,z_4) +C_\ell^{\g\im}(z_1,z_4)C_\ell^{\im \g}(z_2,z_3)\right].  \nonumber
\end{align}
All terms are affected by cosmic variance. In addition, the galaxy-galaxy correlation is affected by shot noise when $z_1=z_3$
\begin{equation}
C_\ell^{\g\g\,{\rm sn} }(z_1,z_3)=\frac{\delta_{z_1,z_3}}{\bar N(z_1)},
\end{equation}
where $\bar N(z_1)$ denotes the mean number of galaxies per redshift bin and steradian. The $\im$-$\im$ correlation is affected by shot noise and interferometer noise; however, it has been shown in~\cite{Shaw:2013wza, Shaw:2014khi} that the former is always subdominant with respect to the latter. In~\cite{Jalilvand2019}, we derived an expression for interferometer noise for \mbox{HIRAX}, based on an analytical expression derived in~\citep{Bull:2014rha}, adapted to the outcome of numerical simulations for HIRAX~\citep{Shaw:2013wza,Shaw:2014khi}. 
 
Finally, the cross-correlation between the galaxy and intensity mapping is not affected by shot noise or interferometer noise~\footnote{Note that, as shown in~\cite{Wolz:2017rlw}, there is a residual shot noise contribution in the cross-correlation due to the galaxies which contribute to both the number counts and the intensity mapping signal. This contribution is however negligible in the covariance, since it adds to the covariance from HI-HI, which is dominated by interferometer noise, that is always significantly larger than shot noise~\cite{Shaw:2013wza, Shaw:2014khi}.}. 

\begin{figure*}
\begin{minipage}{\textwidth}
      \includegraphics[width=0.495\textwidth]{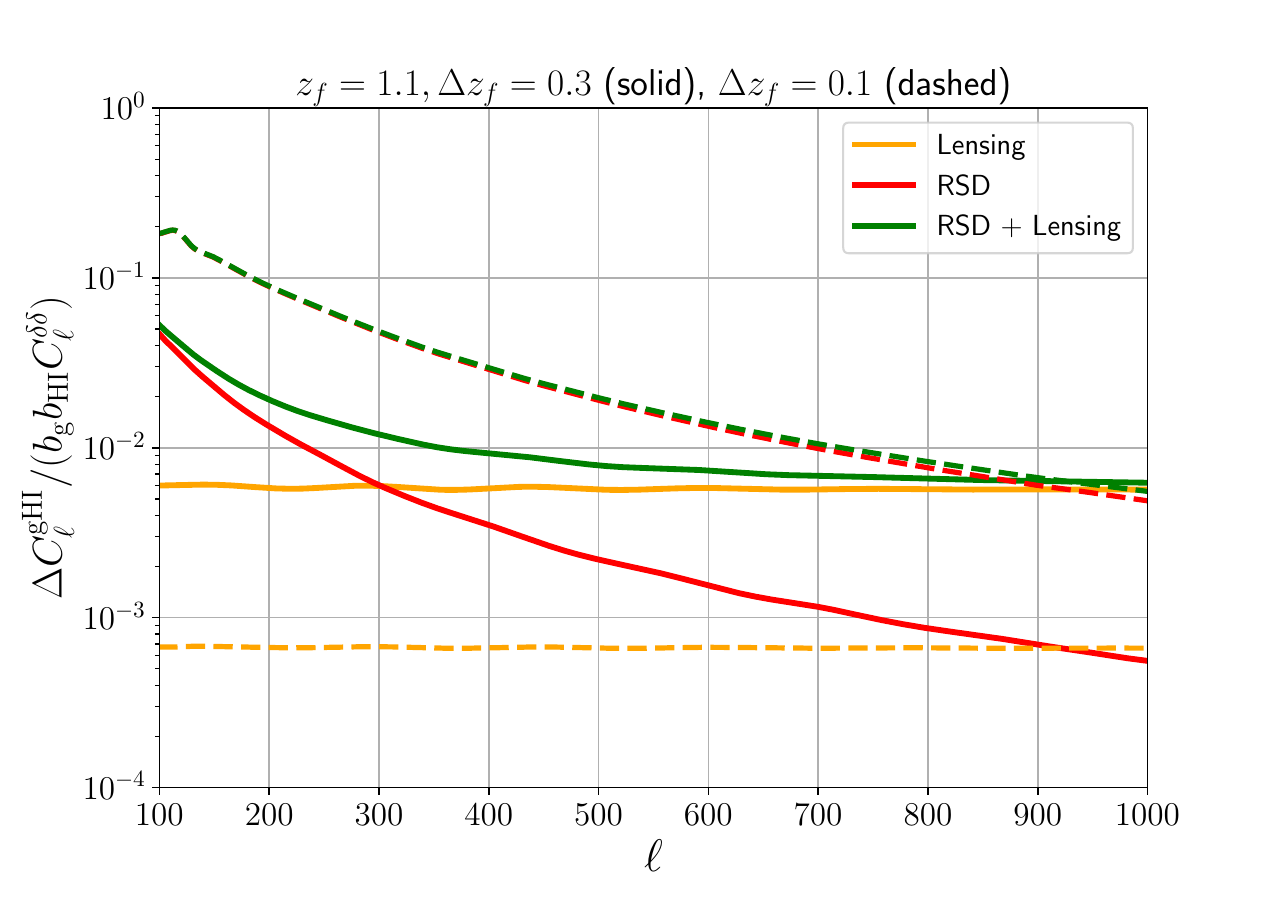}
        \includegraphics[width=0.495\textwidth]{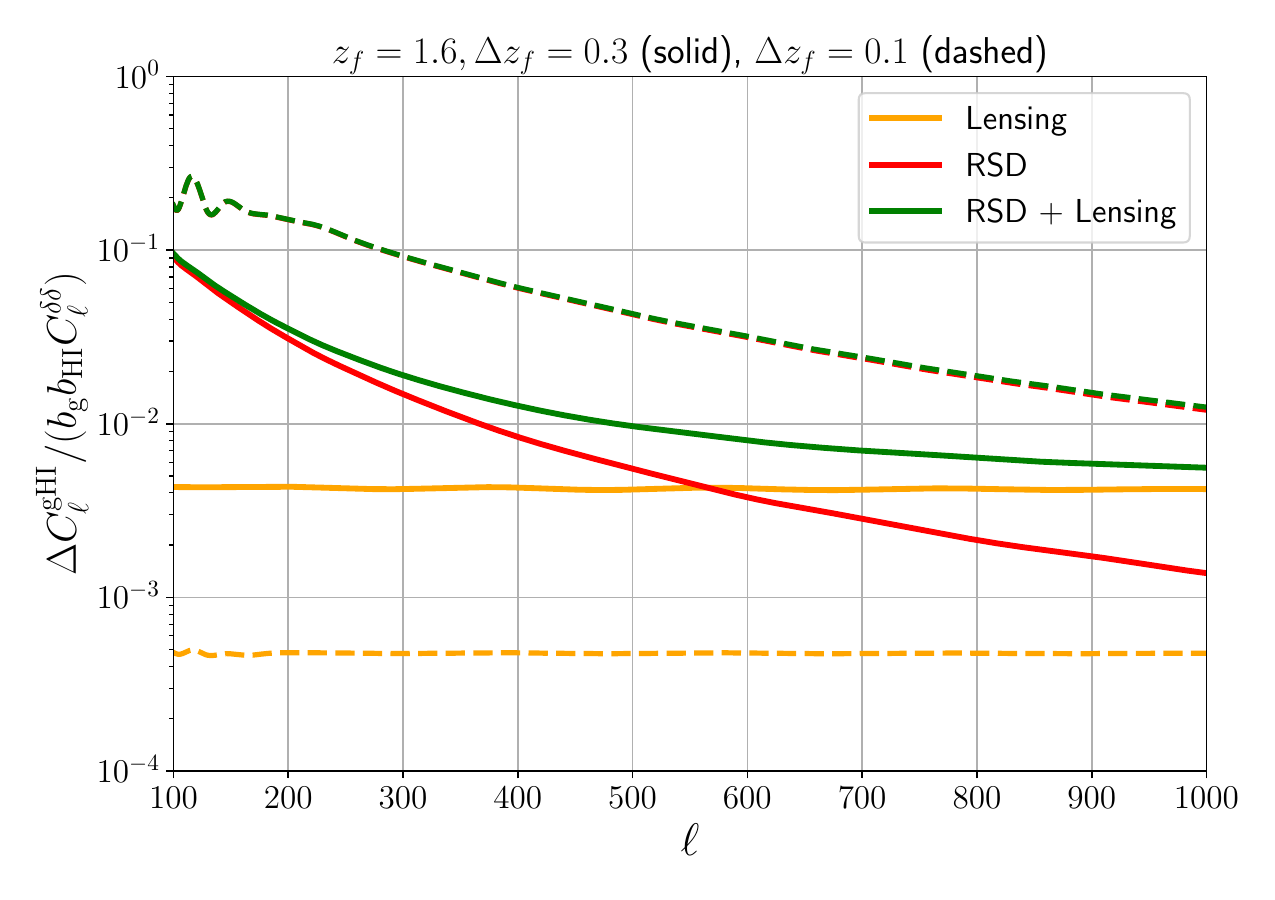}
        \caption{Relative contamination to $C_{\ell}^{\text{g}\im}(z_f,z_f)$ due to lensing magnification and RSD. The left panel shows relative contaminations at $z_f =1.1$ and the right panel at $z_f =1.6$. For each panel there are two sets of curves: $\Delta z_f=0.1$ (dashed) and $\Delta z_f=0.3$ (solid). For the narrower choice of $\Delta z_f$, the RSD contribution dominates, so that the green and red dashed curves nearly coincide. } 
        \label{fig:Cl_cont0.8}
\end{minipage}
\end{figure*}

As shown in Appendix~\ref{app:covariance}, the covariance of $\hEG$ is also affected by the variance of $\beta_\g$. This quantity is measured from the multipoles of the correlation function (or the power spectrum), which are directly sensitive to RSD. Spectroscopic surveys are designed to measure $\beta_\g$ very precisely and are expected to reach a precision of 1\%~\cite{Amendola:2016saw}. Photometric surveys can also measure $\beta_\g$ but with much less precision, of the order of 10\%~\cite{Asorey:2013una}. We use these two values for the spectroscopic and photometric forecasts. Note that, as shown in Appendix~\ref{app:covariance}, the variance of $\beta_\g$ generates non-diagonal contributions, with $\ell\neq\ell'$, to the covariance of $\hEG$. This is because $\beta_\g$ is a parameter measured from the full correlation function in each redshift bin. Its error is, therefore, fully correlated for different values of the angular multipoles $\ell$ and $\ell'$. Neglecting these correlations, as has been done in~\cite{Pourtsidou:2015ksn}, would underestimate the $\beta_\g$ contribution to the variance by a factor of roughly $2\ell+1$.

Finally, the covariance of $\hEG$ depends on the covariance between $C_\ell^{\g\im}$ and $\beta_\g$. Even though this covariance is not precisely zero (since $\beta_\g$ and $C_\ell^{\g\im}$ can be measured from the same volume), it is highly suppressed and can be neglected: as has been shown in~\cite{Taylor:2022rgy}, $C_\ell^{\g\im}$ is mainly insensitive to small radial modes, due to the size of the redshift bins, which washes out small-scale information, whereas $\beta_\g$ is measured from small radial modes. This leads to a negligible covariance between these two types of measurements.

\subsection{Contaminations}
\label{sec:contaminations}

From Eq.~\eqref{eq:sn_signal} we see that the SNR depends on the range of multipoles used in the analysis and the redshift bins. We determine those to reduce the contaminations to $\hEG$.
As discussed previously, the numerator in $\hEG$ is not affected by the lensing contamination that affects standard $\hEG$ estimators~\cite{MoradinezhadDizgah:2016pqy}. However, when $z_f$ is close to $z_b$, the numerator is affected by density-density contamination, given by the last line in Eq.~\eqref{eq:El}. In order to reduce this contamination to acceptable levels, once the redshift bins have been fixed, we will compute the contamination for each pair and remove those for which it is too high. 

\begin{figure*}[t]
    \centering
        \includegraphics[width=0.45\linewidth]{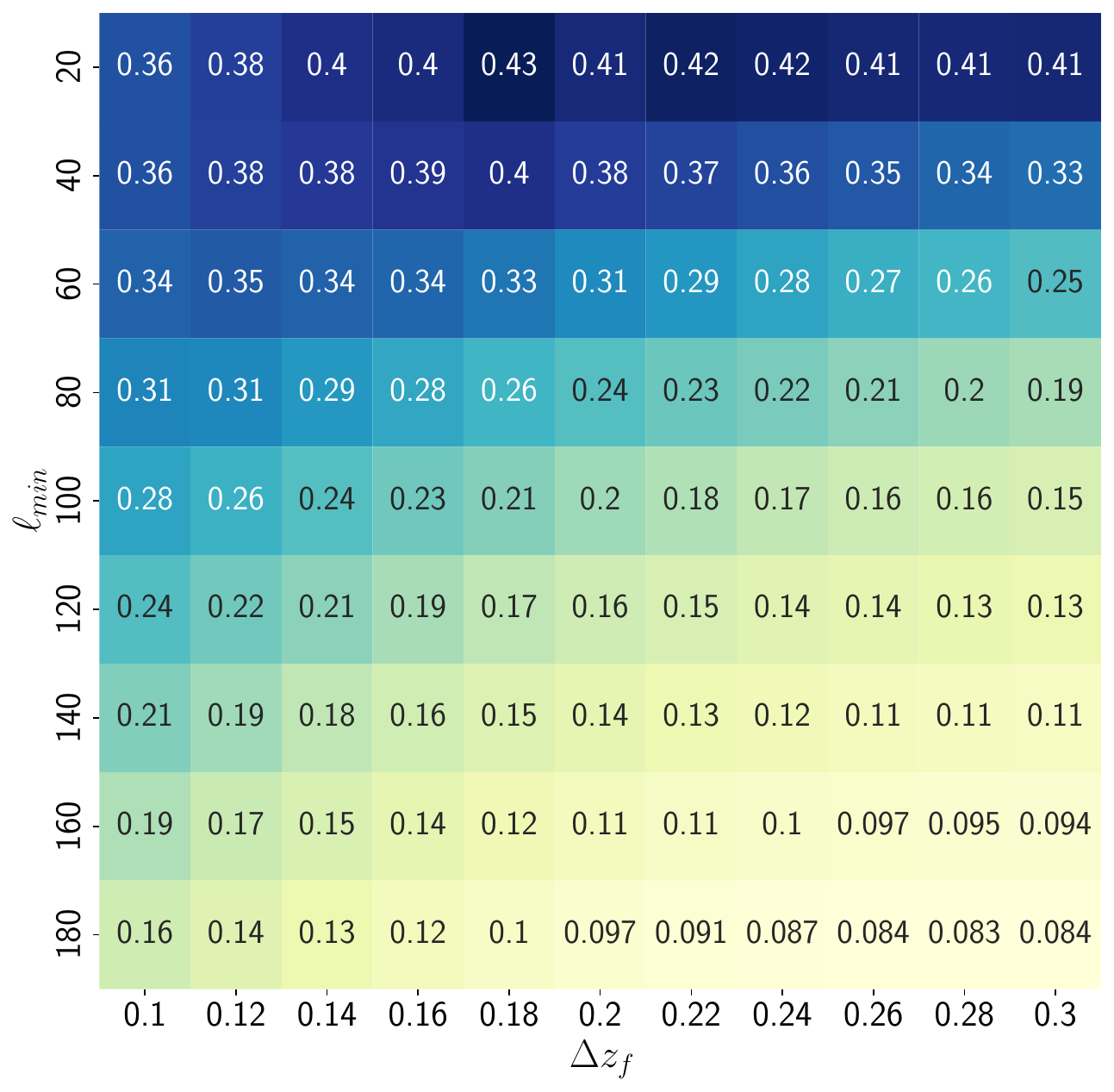}
        \includegraphics[width=0.49\linewidth]{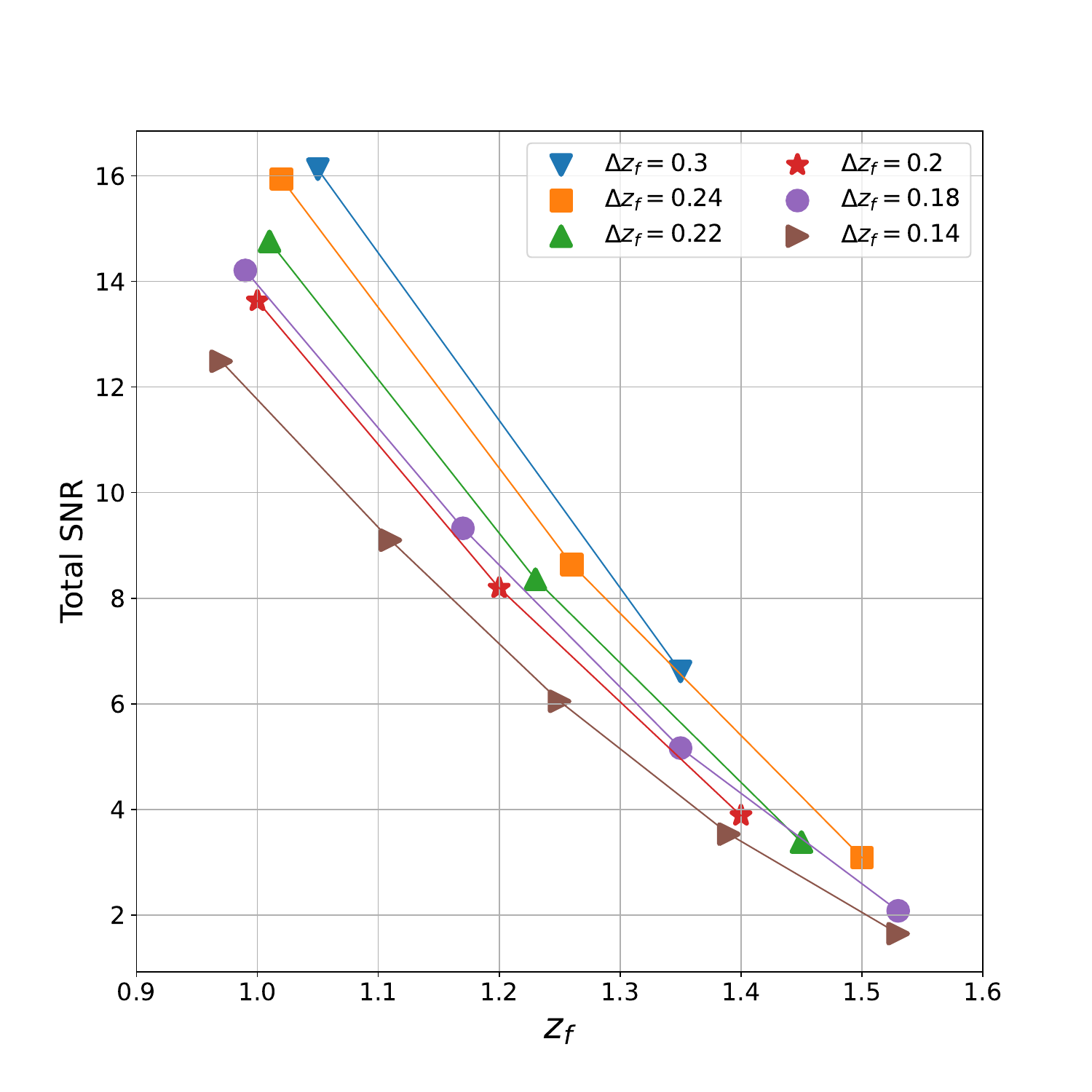}
        \caption{{\it Left panel}: SNR contamination from RSD and lensing magnification for a \emph{Euclid}-like spectroscopic survey as a function of $\ell_{\text{min}}$ and $\Delta z_f$, for $z_f=1.1$, $z_b = 1.6$, and $\Delta z_b = 0.22$. {\it Right panel}: Total SNR as a function of $z_f$ for different values of $\Delta z_f$. Note that in each case, the size of the last redshift bin is adapted to cover the whole available redshift range.}
   \label{fig:SNR_contlmax}
\end{figure*}

The denominator in $\hEG$ is affected by contamination from lensing magnification and by contamination from RSD. The first one has been discussed in~\cite{MoradinezhadDizgah:2016pqy}, whereas the second one is usually ignored.
We aim to choose the widths of the redshift bins to minimize these two contaminations without compromising the SNR. The contribution from RSD and lensing magnification to the galaxy-HI spectrum at equal redshift can be expressed as
\begin{align}
    &\Delta C^{\g\im}_{\ell}(z_f,z_f) = b_\im(z_f)
    \big(5 s(z_f)-2\big)C_{\ell}^{\delta\kappa}(z_f,z_f) \label{eq:cont}\\
    &\!\!\!+ \big[b_\g(z_f)+b_\im(z_f)\big]C_{\ell}^{\delta\,{\rm RSD}}(z_f,z_f)
    + C_{\ell}^{{\rm RSD\,RSD}}(z_f,z_f)\, .\nonumber
\end{align}
The term in the first line is the lensing magnification contamination, and the two terms in the second line are the contaminations from RSD. 
In Fig.~\ref{fig:Cl_cont0.8} we show the relative contamination from lensing magnification and RSD for $z_f=1.1$  and $z_f = 1.6$  and for two values of the width: $\Delta z_f=0.1$ and $\Delta z_f=0.3$. We choose here the \emph{Euclid}-like spectroscopic survey specifications. We see that for thin bins (dashed curves), the contamination due to lensing is less than $0.1\%$ for both redshifts, $z_f =1.1$ and $z_f = 1.6$, while RSD contributes more than 1\% on a wide range of scales. For thick bins (solid curves), the situation is different. Here RSD is still the dominant contamination on large scales, but on small scales ($\ell \gtrsim 300$ for $z_f=1.1$ and $\ell \gtrsim 550$ for $z_f = 1.6$) the lensing magnification contamination dominates. We see that this contamination does not increase with redshift. This is because, in Limber approximation, only the intra-bin lensing contributes. This contribution decreases with redshift roughly at the same rate as the density-density contribution. However, the bias increases faster with redshift than the magnification bias factor $5s-2$ (see Fig.~\ref{fig:gal-mag-bias} in Appendix~\ref{app:surveys}), leading, overall, to a slight decrease with redshift of the relative lensing contamination. Overall the wider redshift bin choice is clearly better in terms of total contamination, which remains below 3\% for $\ell \gtrsim 100$ for $z_f=1.1$ and $\ell \gtrsim 200$ for $z_f = 1.6$. 

Based on Fig.~\ref{fig:Cl_cont0.8} we see that for each foreground redshift $z_f$, we can choose a width $\Delta z_f$ that would minimise the total contamination. Here we have compared the contamination with the density contribution $b_\g b_\im C^{\delta\delta}_{\ell}$. In practice, however, what will determine if our estimator is biased or not is the size of the contamination with respect to the variance: as a rule of thumb, if the contamination leads to a contribution in $\hEG$ which is of the same order as its variance, we expect a bias of the order of 1$\sigma$ on the best-fit parameters extracted from $\hEG$. We define, therefore, the following SNR associated with the contaminations
\begin{align}
\label{eq:sn_contmn}
&\left(\frac{\Delta \rm S}{\rm N}\right)^2(z_f)\equiv\sum_{\ell,\ell'=\ell_{\rm min}}^{\ell_{\rm max}}\sum_{z_b, z_b'=z_{\rm min}}^{z_{\rm max}}\Delta \hat{E}_G(\ell, z_f, z_b)\\
&\times {\rm cov}^{-1} \big[\hat{E}_G(\ell, z_f, z_b),\hat{E}_G(\ell', z_f, z_b')\big]\Delta \hat{E}_G(\ell', z_f, z_b')\, .\nonumber
\end{align}
Here $\Delta \hat{E}_G$ is the difference between the contaminated signal, and the uncontaminated one:
\begin{align}
\Delta &\hat{E}_G (\ell, z_f, z_b)= \\ &\Bigg\{\frac{E_\ell^\times(z_f,z_b)}{\beta_{\g}\left(z_{f}\right) \left(b_\g \left(z_{f}\right)b_\im\left(z_{f}\right) C_\ell^{\delta \delta}\left(z_{f}, z_{f}\right)+\Delta C^{\g\im}_\ell \left(z_{f}, z_{f}\right)\right)}\nonumber\\
&-\frac{E_\ell^\times(z_f,z_b)}{\beta_{\g}\left(z_{f}\right) b_\g(z_f) b_\im(z_f) C_\ell^{\delta \delta} \left(z_{f}, z_{f}\right)}\Bigg\}\frac{1}{\Gamma(z_f,z_b)}\, ,
\end{align}
where $\Delta C^{\g\im}_\ell$ contains the contamination from RSD and lensing, defined in Eq.~\eqref{eq:cont}.

Our strategy to choose optimal redshift bins is based on two factors: minimising the contamination~\eqref{eq:sn_contmn} and maximising the total signal to noise~\eqref{eq:sn_signal}. There are four variables that we need to optimise: $\ell_{\text{min}}$, $\ell_{\text{max}}$, $\Delta z_f$, and $\Delta z_b$. We choose $\ell_{\rm max}=1000$, since above this value, the non-linear modelling of lensing magnification and density fluctuations is uncertain.
In Fig.~\ref{fig:SNR_contlmax} (left panel) we plot the SNR contamination as a function of $\ell_{\rm min}$ and $\Delta z_f$, for the redshift pair $(z_f=1.1, z_b=1.6)$, and for fixed $\Delta z_b=0.22$. We show the results for the case of the \emph{Euclid}-like spectroscopic survey. We see that at fixed $\ell_{\rm min}$, the contamination decreases as we increase the bin size. This is because the RSD contamination decreases for thick bins. The lensing contamination, on the other hand, increases with the size of the bin, but it never overcomes the contamination from RSD once we sum over multipoles. Moreover, the contamination decreases quickly when increasing $\ell_{\rm min}$. This is again due to RSD, which peaks at low $\ell$ as can be seen from Fig.~\ref{fig:Cl_cont0.8}. 

To optimally choose $\Delta z_f$ and $\ell_{\rm min}$, we need to put Fig.~\ref{fig:SNR_contlmax} in balance with the SNR for $\hEG$. Clearly, decreasing $\ell_{\rm min}$ will increase the SNR. A good compromise is to choose $\ell_{\rm min}=100$. This is also motivated by the fact that IM surveys like HIRAX may not have good enough calibration to precisely measure lower multipoles~\cite{Bull:2014rha}. Concerning the size of the bins: the SNR at each $z_f$ increases with the size of the bins, but if we want to test the evolution of $\hEG$ with redshift, we want to have a reasonable number of foreground bins. In Fig.~\ref{fig:SNR_contlmax} (right panel), we plot the SNR summed over all background bins as a function of foreground redshift $z_f$ and for different sizes of the bins. We see that $\Delta z_f =0.24$ is optimal: it allows us to measure $\hEG$ for three foreground redshifts (versus two for $\Delta z_f =0.3$) and has a total SNR of 18.4, larger than for all other cases. The thinner choice, $\Delta z_f=0.14$, is interesting since it allows us to measure $\hEG$ in a higher number of redshift bins. However, the overall SNR (summed over redshift) is 8 percent lower than for $\Delta z_f =0.24$. We therefore choose $\ell_{\rm min}=100$ and $\Delta z_f = 0.24$, which lead to a small SNR contamination of 0.17. Note that if we reduce $\ell_{\rm min}$ to 60, the SNR increases only very slightly from 18.4 to 18.9, whereas the contamination increases from 0.17 to 0.28. We conclude therefore that $\ell_{\rm min}=100$ is a better choice. This procedure needs to be performed for each value of the foreground redshift $z_f$. We find similar results at all redshifts, with a maximum SNR contamination of 0.21 for the pair $z_f=1.02$, and $z_b=1.71$, which is still perfectly acceptable. 

We then repeat the same procedure for the \emph{Euclid}-like photometric and the SKA2-like surveys. In Table \ref{tab:bins}, we list the optimal binning for each case.

\begin{table}[t]
\centering
\caption{Optimal redshift bin configurations for \emph{Euclid}-like spectroscopic, photometric and SKA2-like surveys.}
\label{tab:bins}
\begin{tabular}{C{1cm}  C{1cm} C{1cm}  C{1cm} C{1cm}  C{1cm}}
  \toprule 
  \multicolumn{2}{c}{\emph{Euclid} spectro} & \multicolumn{2}{c}{\emph{Euclid} photo} & \multicolumn{2}{c}{SKA2} \\
  \cmidrule(rl){1-2} \cmidrule(rl){3-4} \cmidrule(rl){5-6}\cmidrule(rl){5-6}
  $z$    &  $\Delta z$ & $z$ & $\Delta z$& $z$&$\Delta z$  \\
  \midrule
 1.02    &  0.24 & 0.84 & 0.2& 1.02& 0.24  \\
  
 1.26    &  0.24 & 1.00 & 0.2& 1.26& 0.24  \\
  
 1.50    &  0.24 & 1.14 & 0.2& 1.50& 0.24  \\
 
 1.71    &  0.18 & 1.30 & 0.2& 1.74& 0.24  \\
 
      &   & 1.44 & 0.2& 1.93& 0.14  \\
   
   &   & 1.62 & 0.2& &  \\ 
      &   & 1.78 & 0.5& &  \\ 
         &   & 1.91 & 0.5& &  \\ 
   
  \bottomrule
\end{tabular}
\end{table}

\begin{table}[t]
\centering
\caption{SNR contamination from density for the \emph{Euclid}-like spectroscopic survey.}
\label{table:densSNRspec}
\begin{tabular}{c C{1.6cm} C{1.6cm} C{1.6cm}}
\toprule
$z_f$  & $z_b = 1.26$ & $z_b = 1.50$ & $z_b = 1.71$ \\ [0.5ex] % inserts table %heading
\midrule
$1.02$   & 0.09&  $3 \times 10^{-16}$ & $7 \times 10^{-16}$ \\
$1.26$   & &  $0.006$ & $4 \times 10^{-16}$ \\
$1.50$   & &   & $0.04$ \\
\bottomrule
\end{tabular}
\end{table}

\begin{table}[t]
\centering
\caption{SNR contamination from density for the SKA2-like survey.}
\label{table:densSNR-SKA}
\begin{tabular}{c C{1.6cm} C{1.6cm} C{1.6cm} C{1.6cm}}
\toprule
$z_f$  & $z_b = 1.26$ & $z_b = 1.50$ & $z_b = 1.74$ & $z_b = 1.91$\\ [0.5ex] % inserts table %heading
\midrule
$1.02$   & 1.40&  $10^{-16}$ & $10^{-16}$ & $ 10^{-16}$\\
$1.26$   & &  $0.94$ & $10^{-16}$ &$ 10^{-16}$ \\
$1.50$   & &   & $0.60$ & $10^{-17}$\\
$1.74$   & &   &  & $1.54$\\
\bottomrule
\end{tabular}
\end{table}

\begin{table}[t]
\centering
\caption{SNR contamination from density for the \emph{Euclid}-like photometric survey.}
\label{table:densSNR-photo}
\begin{tabular}{c C{1.0cm} C{1.0cm} C{1.0cm} C{1.0cm} C{1.2cm} C{1.0cm} C{1.0cm}}
\toprule
$z_f$ & $z_b = 1.0$ & $z_b = 1.14$ & $z_b = 1.30$ & $z_b = 1.44$ & $z_b = 1.62$ & $z_b = 1.78$ &  $z_b = 1.91$\\ [0.5ex] % inserts table %heading
\midrule
$0.84$ & $40.8$ & $3.7$  & $0.25$& $ 0.004$ & $ 10^{-6}$ & $ 10^{-3}$ & $10^{-3}$ \\
$1.00$ & &   $22.6$ & $1.8$& $0.15$& $ 10^{-3}$ & $ 0.1$ & $0.03$\\
$1.14$ &  & & $27.1$& $4.3$ & $0.08$ & $ 0.24$ & $0.14$\\
$1.30$ & &   & & $12.8$ & $1.4$ & $2.7$ & $1.3$\\
$1.44$ &  &   & &  & $5.7$ & $10.3$ & $5.8$\\
$1.62$ &  &   & &  &  & $66.8$& $51.8$\\
$1.78$ &  &   & &  &  & &$28.7$ \\
\bottomrule
\end{tabular}
\end{table}

Once the binning has been fixed, we compute the SNR contamination from density in the numerator of $\hEG$ (due to the last line in Eq.~\eqref{eq:El}), for each pair $(z_f, z_b)$. The results for the \emph{Euclid}-like spectroscopic survey are shown in Table \ref{table:densSNRspec}. This SNR contamination is at most 9\% (for neighbouring pairs). We can therefore keep all pairs in our forecast. For SKA2, the density contamination for adjacent pairs is more significant, as seen in Table~\ref{table:densSNR-SKA}. This is because the variance of $\hEG$ is smaller for SKA2 due to the broader sky coverage and lower shot noise. Consequently, the contamination from density relative to the variance is larger for SKA2 than for \emph{Euclid} spectroscopic. To avoid bias in our estimator, we remove adjacent pairs for SKA2.

For \emph{Euclid}-like photometric, the situation is different since we are using Gaussian window functions. In this case, there is still a significant overlap of non-consecutive bins, especially at high redshift, leading to non-negligible density contamination, as seen from Table~\ref{table:densSNR-photo}. Based on these results, we remove all pairs with contamination larger than 0.25. With this, we ensure that the bias from such contamination is below 0.25$\sigma$, which is acceptable. We see that with this criteria, we have only three foreground bins for \emph{Euclid}-like photometric. Note that in reality, we expect the contamination to be smaller: here, we have used a Gaussian window for HIRAX as well, due to the limitation of {\sc class} to have two different window functions. In practice, however, HIRAX can select bins with sharp edges for which the bins overlap, and consequently, the density contamination will be reduced.

Overall, our choice of binning is such that the total contamination from RSD, lensing and density is at most 0.3$\sigma$ for \emph{Euclid}-like spectroscopic (for $z_f=1.02$), 0.6$\sigma$ for SKA2-like (for $z_f=1.02$), and 0.4$\sigma$ for Euclid-like photometric (for $z_f=1.14$). This means that our estimator is robust and will lead to a measurement of $\Omega_{\rm m} \Sigma/f$ that is not significantly biased (half a sigma at most). Note that one possibility to reduce the bias further would be to model the contamination using a reference cosmology, and to add it to the model. This `reference contamination' can then be kept fixed. This is not correct (since the contamination depends on cosmology), but since it is already subdominant, the bias generated by keeping it fixed would be very small.

\section{Modified gravity forecasts}
\label{sec:forecasts}

From Eq.~\eqref{eq:EGhat}, we see that constraints on $\hEG(z_f)$ directly translate into constraints on the combination of parameters $\Omega_{\rm m}\Sigma(z_f)/f(z_f)$. For each $z_f$, the Fisher element for this combination reads
\begin{align}
\mathcal{F}&=\sum_{\ell,\ell'=\ell_{\rm min}}^{\ell_{\rm max}}\sum_{z_b, z_b'=z_{\rm min}}^{z_{\rm max}}\frac{\text{d} \hat{E}_G(\ell, z_f, z_b)}{\text{d} (\Omega_{\text{m}}\Sigma/f)}\\
&\times {\rm cov}^{-1} \big[\hat{E}_G(\ell, z_f, z_b),\hat{E}_G(\ell', z_f, z_b')\big]\frac{\text{d} \hat{E}_G(\ell', z_f, z_b')}{\text{d} (\Omega_{\text{m}}\Sigma/f)}\nonumber\\
&=\sum_{\ell,\ell'=\ell_{\rm min}}^{\ell_{\rm max}}\sum_{z_b, z_b'=z_{\rm min}}^{z_{\rm max}}\!\!{\rm cov}^{-1} \big[\hat{E}_G(\ell, z_f, z_b),\hat{E}_G(\ell', z_f, z_b')\big]\, .\nonumber
\label{eq:fisher}
\end{align}

\begin{figure}
\begin{minipage}{.48\textwidth}
    \centering
    \includegraphics[width=1.0\linewidth]{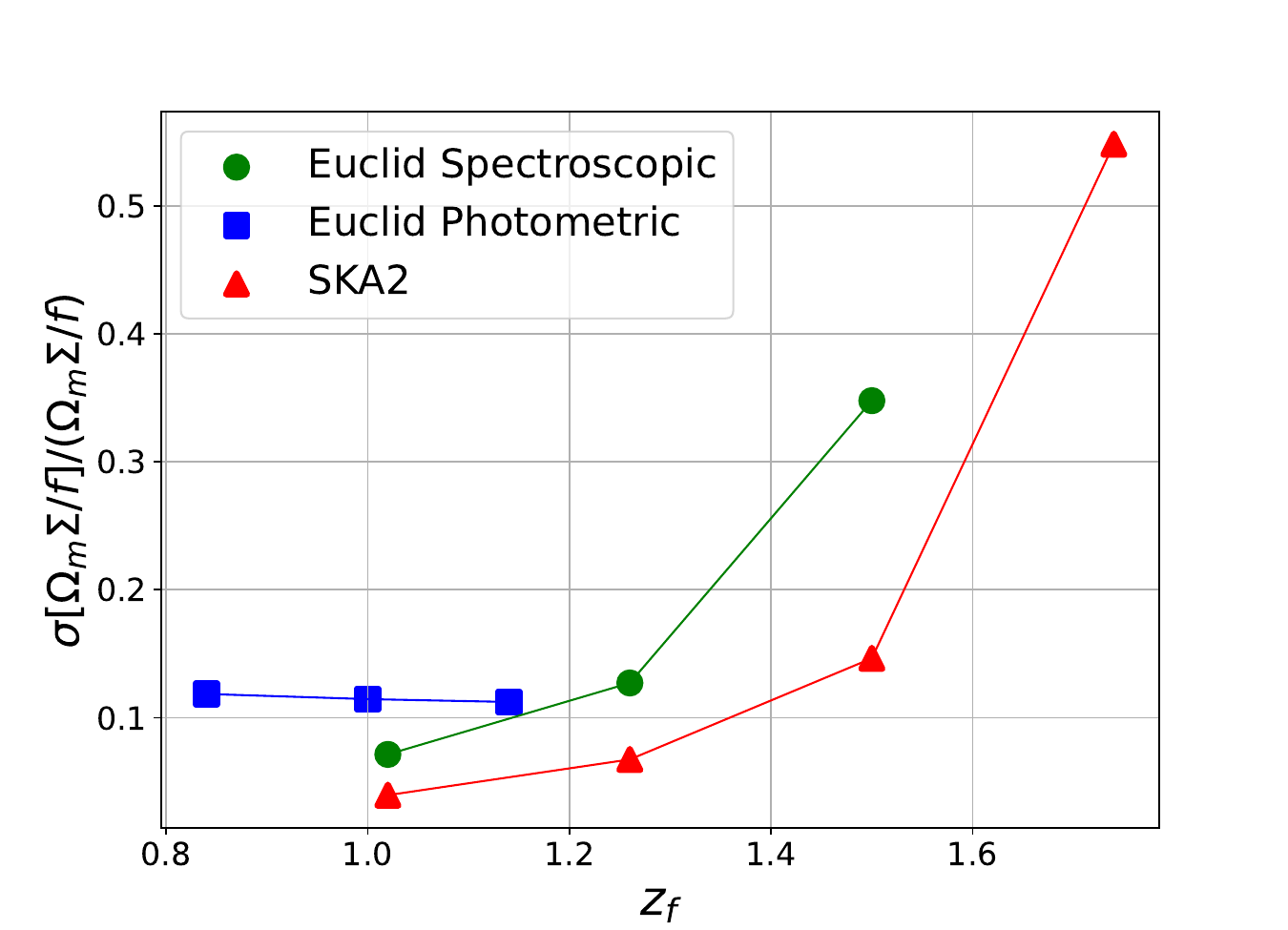}
    \caption{Relative uncertainty on $\Omega_{\text{m}}\Sigma/f$ as function of foreground redshift for \emph{Euclid}-like spectroscopic, \emph{Euclid}-like photometric and SKA2-like. In the Fisher analysis we use $\ell_{\text{min}}=100$, $\ell_{\text{max}}=1000$, and the redshift bins given in Table~\ref{tab:bins}. }
    \label{fig:Fisher}
\end{minipage}
\end{figure}

In Fig.~\ref{fig:Fisher} we show the relative error on $\Omega_{\rm m}\Sigma/f$ for the three combinations of surveys, using the optimal redshift bins defined in Table.~\ref{tab:bins}. The constraints expected from a \emph{Euclid}-like spectroscopic survey are very good at low redshift, around 7\%. The precision decreases as the foreground redshift increases, reaching 32\% at $z_f=1.50$. The fact that the constraints are significantly better at low redshift is due to two reasons: firstly, when $z_f$ is small, we can correlate it with a larger number of background redshift bins, which increases the precision of the measurement. Secondly, at low redshift, the number density for a \emph{Euclid}-like spectroscopic survey is significantly larger than at high redshift, leading to a strong suppression of shot noise.

On the other hand, a \emph{Euclid}-like photometric survey has a precision of $\sim 11\%$ for all foreground redshifts. The primary source of error for photometric surveys is the error on $\beta_\g$ (of 10\%) since this parameter cannot be well constrained from a photometric survey. This consequently limits the precision with which $\hEG$ can be measured. To model this uncertainty correctly, it is crucial to include the $\beta_\g$ covariance fully correlated in $\ell$ as in \eqref{eq:covarianceEG}. As shown in Appendix~\ref{app:covariance}, this non-diagonal contribution follows from the fact that $\beta_\g$ is measured by combining all available scales (from the multipoles of the power spectrum) and not independently for each value of $\ell$. Neglecting these non-diagonal terms in the covariance leads to an uncertainty on $\hEG$ that is smaller than the uncertainty on $\beta_g$ alone, like for example in \cite{Pourtsidou:2015ksn}, where $\hEG$ is predicted to be measured with a precision of 1\% with LSST, even though LSST can measure $\beta_\g$ with a precision of 10\% only, which is not possible. Similarly, in~\cite{Pullen:2014fva}, $\hEG$ is forecasted to be measured with a precision of 1\% with LSST and photometric \emph{Euclid}, using CMB lensing, assuming that $\beta_\g$ can be measured with a 10\% uncertainty. Again, this is not possible: $\hEG$ cannot be measured better than $\beta_\g$ is. As before, the error in these forecasts comes from the fact that the error on $\beta_\g$ is wrongly assumed to be uncorrelated in $\ell$ (see their Eq. (17).

The constraints predicted for SKA2 are tighter by a factor of 2 with respect to the constraints from \emph{Euclid}-like spectroscopic, reaching $3.9\%$ at low redshift. This is mainly due to the higher number density of SKA2, the broader sky coverage, and the larger redshift range. However, the constraints are quite poor in the last redshift bin, of 54\% only. This is due to the large shot noise at high redshift, which significantly degrades the constraints, and to the fact that only one pair can be used for this case.

\begin{figure}
    \centering
    \includegraphics[width=0.49\textwidth]{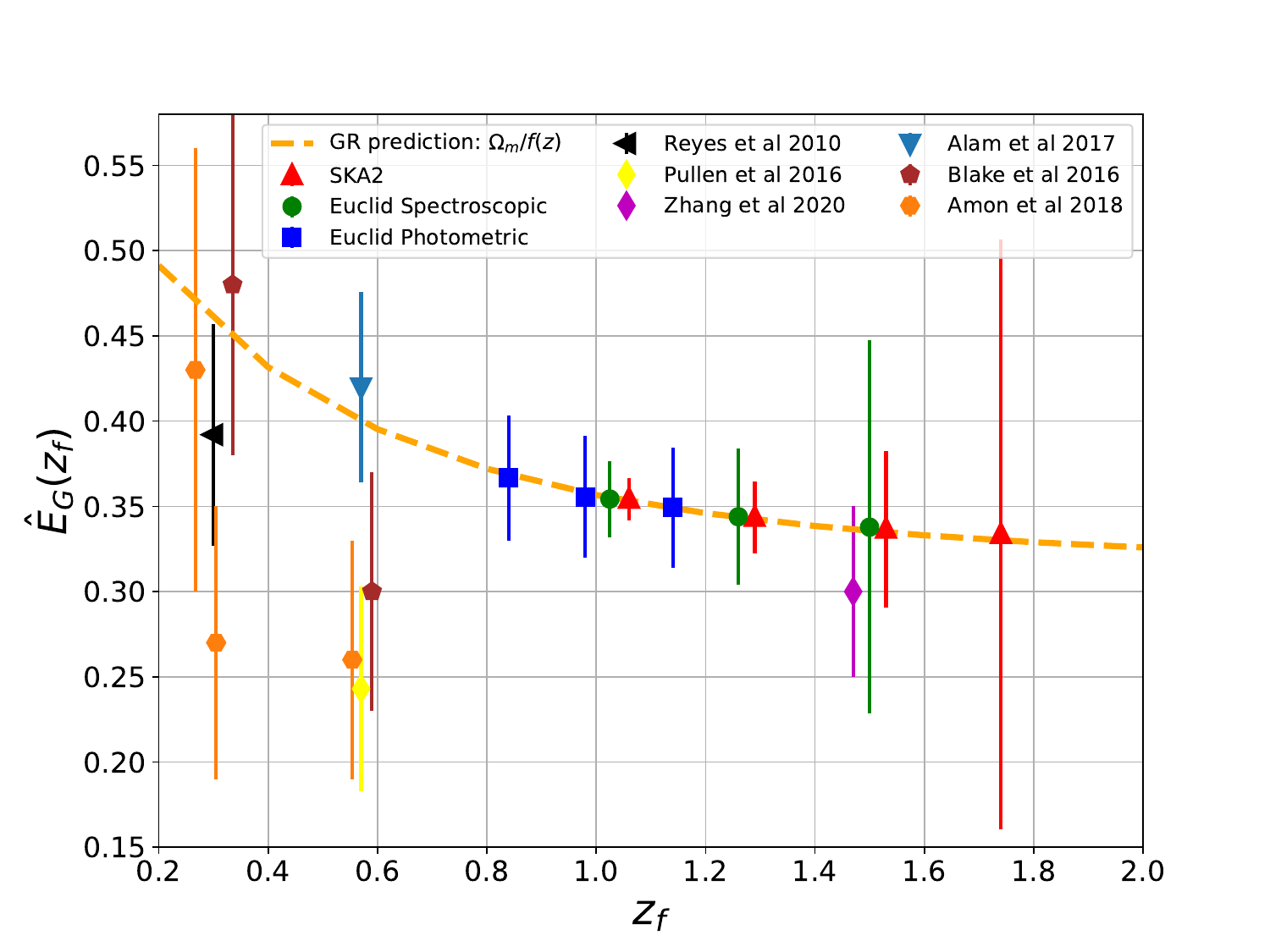}
    \caption{$\hEG$ and its error bars, as function of foreground redshift, $z_f$. We compare our results from Fisher forecasts with measurements from~\cite{Reyes:2010tr, Pullen:2015vtb, Amon2018MNRAS.479.3422A, Blake2015, Alam:2016qcl,Zhang:2020vru}. The dashed orange line is the GR prediction, $\hat{E}_{G}(z)= \Omega_{{\text{m}}}/f(z)$ calculated with our fiducial cosmology.}
    \label{fig:Eg-mu-sigma}
\end{figure}

In Fig.~\ref{fig:Eg-mu-sigma}, we compare our forecasted constraints with current measurements of $\hEG$ from~\citep{Reyes:2010tr, Pullen:2015vtb, Amon2018MNRAS.479.3422A, Blake2015}. The orange dashed line represents the GR prediction, $\hat{E}_G(z) = \Omega_{{\text{m}}}/f(z)$, computed with our fiducial choice of cosmology\footnote{Note that the measurements depend on the choice of background cosmology through $\Gamma$ and can therefore not directly be compared with our prediction to assess if there is a tension or not. In Table~\ref{table:eg} we provide for this reason a summary of the tension (or agreement) quoted from the respective studies.}. 
The best measurement for $\hEG$ is the one from~\citep{Alam:2016qcl} at $z_f=0.57$ which has a precision of 13\%. We see from Fig.~\ref{fig:Eg-mu-sigma} that our method to measure $\hEG$ from clustering of galaxies and intensity mapping has the potential to significantly tighten these constraints over a wide range of redshifts, allowing us to test the validity of GR robustly on cosmological scales.

\begin{table}[t]
\centering
\caption{Previous measurements of $\hat{E}_{G}$.}
\label{table:eg}
\begin{tabular}{c c c c}
\toprule
Authors & $z$ & $\hat{E}_G(z)$& Agreement \\ 
\midrule
Reyes et al \citep{Reyes:2010tr}  & $0.37$& $0.392 \pm 0.065$& agrees with GR\\
\hline
\multirow{2}{*}{Blake et al \citep{Blake2016MNRAS.462.4240B} } & \Tstrut$0.32$ & $0.48\pm 0.10$ & agrees with GR  \\
& $0.57$ & $0.30 \pm 0.07$ & 1.4$\sigma$ deviation   \\ 
\hline
\multirow{3}{*}{Amon et al \citep{Amon2018MNRAS.479.3422A} } & \Tstrut$0.27$ & $0.43\pm 0.13$ & agrees with GR   \\
& $0.31$ & $0.27 \pm 0.08$ & $2.3\sigma$ deviation  \\ 
& $0.55$ & $0.26 \pm 0.07$ & $2 \sigma$ deviation  \\ 
\hline
Alam et al \citep{Alam:2016qcl}  & \Tstrut$0.57$& $0.42 \pm 0.056$ & agrees with GR \\
\hline
Pullen et al \citep{Pullen2016}  & \Tstrut$0.57$& $0.24 \pm 0.060$ & $2.6\sigma$ deviation\\
\bottomrule
\end{tabular}
\end{table}
%========================
\section{Conclusion}
\label{sec:conclusion}

Testing the consistency of GR at cosmological scales is one of the main goals of modern cosmology. Since a large number of theories have been constructed where gravity is modified or where a dark energy component has been added, confronting each theory individually with observations is no longer feasible. It is, therefore, necessary to build model-independent tests. In this paper, we focused on one of these tests: the $\hEG$ statistic, which compares the evolution of the sum of the potentials with that of the velocity.

The $\hEG$ statistic has been successfully measured with various data sets, showing mild tensions with the GR predictions in some cases. It is therefore of great interest to see if future surveys will confirm or not these tensions. However, the currently used estimators of $\hEG$ suffer from two problems: first, they are affected by lensing magnification, a contamination that will become important for future surveys and may invalidate their use. Second, they mix different observables (clustering and shear, or clustering and CMB lensing), making them sensitive to tensions between these different data sets, due for example to different systematics.

In this paper, we have proposed an alternative estimator for $\hEG$, which relies {\it only on clustering measurements} from two different tracers: galaxies and intensity mapping. We have shown that this estimator is very robust: it only uses cross-correlations that are less affected by systematics, and by choosing the binning in which $\hEG$ is measured appropriately, the contaminations are below 0.3$\sigma$ for a \emph{Euclid}-like spectroscopic survey, below 0.6$\sigma$ for a SKA2-like survey, and below 0.4$\sigma$ for a \emph{Euclid}-like photometric survey. Moreover, our estimator depends on a choice of fiducial cosmology only through background parameters, namely through $H(z)$ and distances. Contrary to some estimators used in the literature, it does not depend on the evolution of perturbations, like in~\cite{Pullen:2015vtb}.

Using Fisher forecasts, we have found that $\hEG$, and consequently deviations from GR, can be measured over a wide range of redshifts and with a precision of up to 7\% with a \emph{Euclid}-like spectroscopic survey and 3.9\% with a SKA2-like survey. We have also explicitly shown that a \emph{Euclid}-like photometric survey cannot reach such a good precision on $\hEG$. The reason is that for a photometric survey, $\beta_\g$ is measured at best with a precision of 10\%, which intrinsically limits the precision on $\hEG$. This point was not properly accounted for in some previous forecasts, which claim a 1\% precision on $\hEG$ using photometric surveys like \emph{Euclid} or LSST~\cite{Pullen:2014fva,Pourtsidou:2015ksn}. The error in these forecasts comes from the covariance matrix, which does not properly include correlations in $\ell$'s generated by the fact that $\beta_\g$ is measured from a combination of all scales, and not independently for each values of $\ell$.

Our study shows therefore that the combination of intensity mapping and galaxy clustering provides an excellent  strategy to measure $\hEG$ robustly and precisely. This will dramatically improve current constraints and extend them to higher redshifts, allowing us to test deviations from GR very accurately.

\section{Acknowledgments}

It is a pleasure to thank Patrick Simon for useful comments on the manuscript.
MA and CB acknowledge financial support from the European Research Council (ERC) under the European Union’s Horizon 2020 research and innovation program (Grant agreement No.~863929; project title ``Testing the law of gravity with novel large-scale structure observables"). MJ and MK acknowledge financial support from the Swiss National Science Foundation.

%========================

\appendix

\section{Covariance of $\hEG$}
\label{app:covariance}

We compute the covariance of $\hEG(\ell,z_f,z_b)$ for a fixed value of $z_f$ since we measure it independently in each foreground redshift bin. However, since we combine different background redshift bins in the measurements, we need to account for the covariance of $\hEG$ between different background redshifts $z_b$ and $z_b'$. Moreover, since $\beta_\g$, which enters into $\hEG$, is measured by combining all scales (and not independently for each value of $\ell$), it will induce correlations between $\hEG$ at different values of $\ell$.

In general, for $X=f(A,B)$ and $X'=f(A',B')$, where $f$ is a generic function, the covariance of $X$ and $X'$ can be expressed as:
\begin{align}
    \operatorname{Cov}(X,X') =  \sum_{\substack{\alpha= A,B \\ \beta=A',B'}} \frac{\partial X}{\partial \alpha} \operatorname{Cov}(\alpha,\beta) \frac{\partial X'}{\partial \beta}\, .
\end{align}
Using this formula, the covariance of $\hEG$ becomes
\begin{align}
\operatorname{Cov}&[\hEG(\ell,z_f,z_b),\hEG(\ell',z_f,z_b')] =\label{eq:covarianceEG}\\
&\Bigg\{ \frac{\text{Cov}\left[\ec(z_f,z_b),\ec(z_f,z_b')\right]}{\ec(z_f,z_b)\hec(z_f,z_b')} \nonumber\\
     &\quad+ \frac{\operatorname{Var}\left[\Cgh(z_f,z_f)\right]}{\left(\Cgh(z_f,z_f)\right)^2} \nonumber \\
     &\quad- \frac{\operatorname{Cov}\left[\ec(z_f,z_b),\Cgh(z_f,z_f)\right]}{\ec(z_f,z_b)\Cgh(z_f,z_f)} \nonumber \\
    & \quad- \frac{\operatorname{Cov}\left[\ec(z_f,z_b'),\Cgh(z_f,z_f)\right]}{\ec(z_f,z_b')\Cgh(z_f,z_f)} \nonumber \Bigg\}\\
    &\times E_{G}(\ell,z_f,z_b)E_{G}(\ell,z_f,z_b')\,\delta^{\text{D}}_{\ell,\ell'}\nonumber \\
    & + \hEG(\ell,z_f,z_b) \hEG(\ell',z_f,z_b') \frac{\operatorname{Var}\left[\beta_{\text{g}}\right]}{\beta_{\text{g}}^2}\, ,\nonumber
    \label{eq:covEq}
\end{align}
where $\delta^{\text{D}}_{\ell,\ell'}$ represents the Dirac delta function. Here, we have neglected the covariance between $\beta_\g$ and both $C_\ell^{\g\im}$ and $\ec$, as discussed in the main text. 

The covariance of GIMCO, which enters in the second line of Eq.~\eqref{eq:covarianceEG} is given by
\begin{align}
\text{Cov}\Big[&E^{\times}_{\ell}(z_f,z_b),E^{\times}_{\ell}(z_f,z_b')\Big]= \\
&\text{Cov}\left[ C_{\ell}^{\g\text{HI}}(z_b,z_f), C_{\ell}^{\g\text{HI}}(z_b',z_f) \right]\nonumber\\
&- \text{Cov}\left[ C_{\ell}^{\g\text{HI}}(z_b,z_f), C_{\ell}^{\g\text{HI}}(z_f,z_b')\right]\nonumber\\
&- \text{Cov}\left[ C_{\ell}^{\g\text{HI}}(z_f,z_b), C_{\ell}^{\g\text{HI}}(z_b',z_f)\right]\nonumber\\
& +\text{Cov}\left[ C_{\ell}^{\g\text{HI}}(z_f,z_b), C_{\ell}^{\g\text{HI}}(z_f,z_b') \right]\, ,\nonumber
\end{align}
where the covariance of $C_\ell^{\g\im}$ is given in Eq.~\eqref{eq:covClgim}.

Similarly, the covariance of $\ec$ with $C_\ell^{\g\im}$, which enters in the 4th and 5th lines of Eq.~\eqref{eq:covarianceEG} can be written as
\begin{align}
\operatorname{Cov}\Big[&\hec(z_f,z_b), \Cgh(z_f,z_f)\Big]=\\
&\text{Cov}\left[ C_{\ell}^{\g\text{HI}}(z_b,z_f), C_{\ell}^{\g\text{HI}}(z_f,z_f) \right]\nonumber\\
&- \text{Cov}\left[ C_{\ell}^{\g\text{HI}}(z_f,z_b), C_{\ell}^{\g\text{HI}}(z_f,z_f)\right]\nonumber\, .
\end{align}

\section{Galaxy and Intensity Mapping Surveys}
\label{app:surveys}
This paper uses three example surveys for galaxy clustering: \emph{Euclid}-like spectroscopic, \emph{Euclid}-like photometric, and SKA2-like; and one intensity mapping survey: HIRAX. 

The bias of the intensity mapping from HIRAX is given by \citep{Jalilvand2019}
\begin{equation}
    b_{\text{HI}}(z) = 0.677\Big(1 +  3.8 \times 10^{-1} z + 6.7 \times 10^{-2} z^2\Big)\, .
    \label{eq:bIM}
\end{equation}

For the \emph{Euclid}-like photometric survey, we use the bias and magnification bias that have been measured from the flagship simulation in each of the redshift bins we are using, see Table 1 of~\citep{Euclid:2021rez}.

For the SKA2-like survey, we use the bias, and magnification bias modelled in~\cite{Villa:2017yfg}:
\begin{align}
    b_{\text{g}}(z) &= 0.5887 \exp(0.813 z)\, , \\
    s(z) &= s_0 + s_1 z + s_2 z^2 + s_3 z^3\, ,
    \label{eq:bias_magSKA2}
\end{align}
with $s_0 = -0.1068$, $s_1=1.359$, $s_2=-0.620008$, and $s_3 = 0.1885$.

Finally, for the \emph{Euclid}-like spectroscopic survey we use the bias given in~\citep{Euclid:2019clj} (Table 3). Since we are not using the same bins as there, we interpolate this bias with the following function
\begin{equation}
b_\g(z) = -1.094 + 1.813 \sqrt{1+z}\, .\label{eq:Euclidspectro}
\end{equation}
The magnification bias for \emph{Euclid}-like spectroscopic has never been modelled nor measured in simulations. We use, therefore, for simplicity, the expression given in~\citep{Montanari:2015rga}  (which is another model for SKA2)
\begin{equation}
    s(z) = 0.9329 - 1.562\exp(-2.437 z)\, .
    \label{eq:mag_Euclidspec}
\end{equation}

\begin{figure*}
    \centering
    \includegraphics[width=0.47\textwidth]{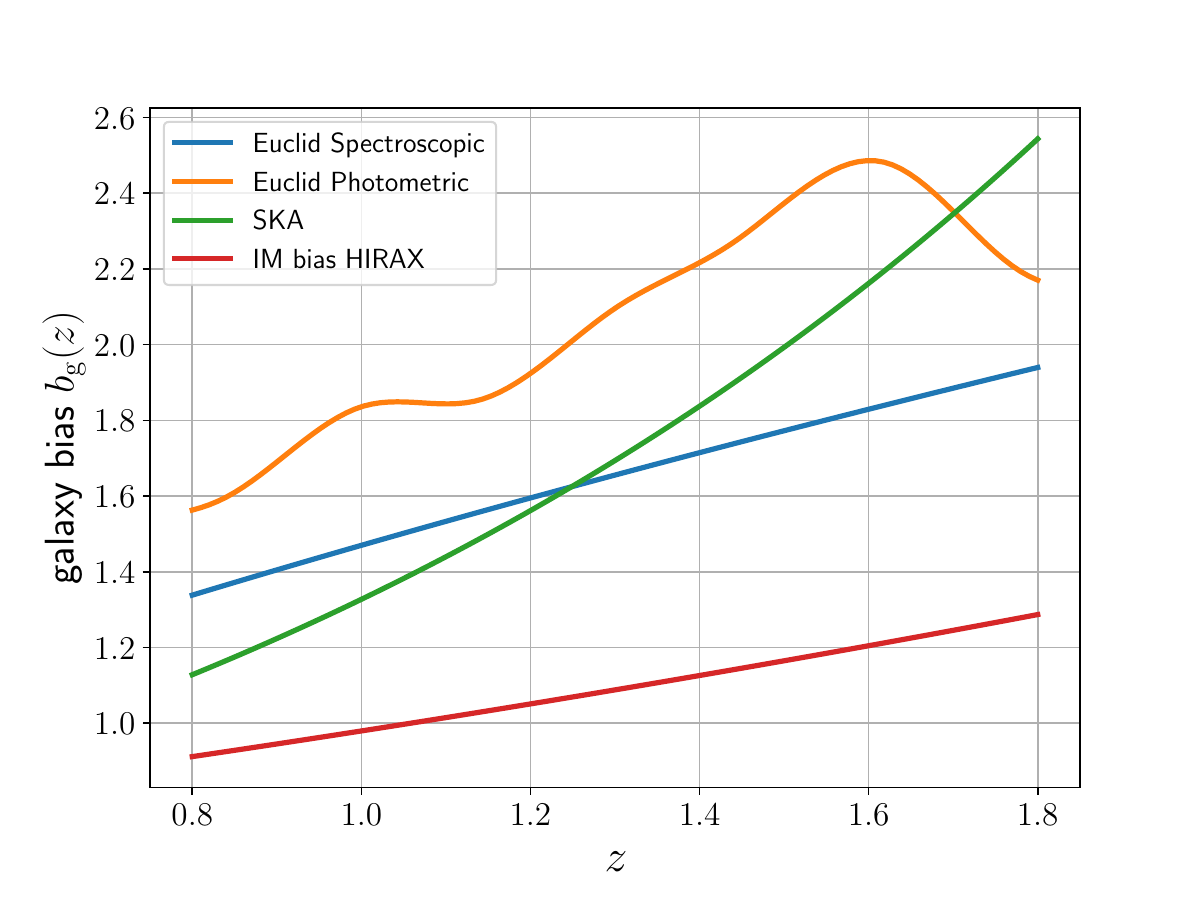}
    \includegraphics[width=0.47\textwidth]{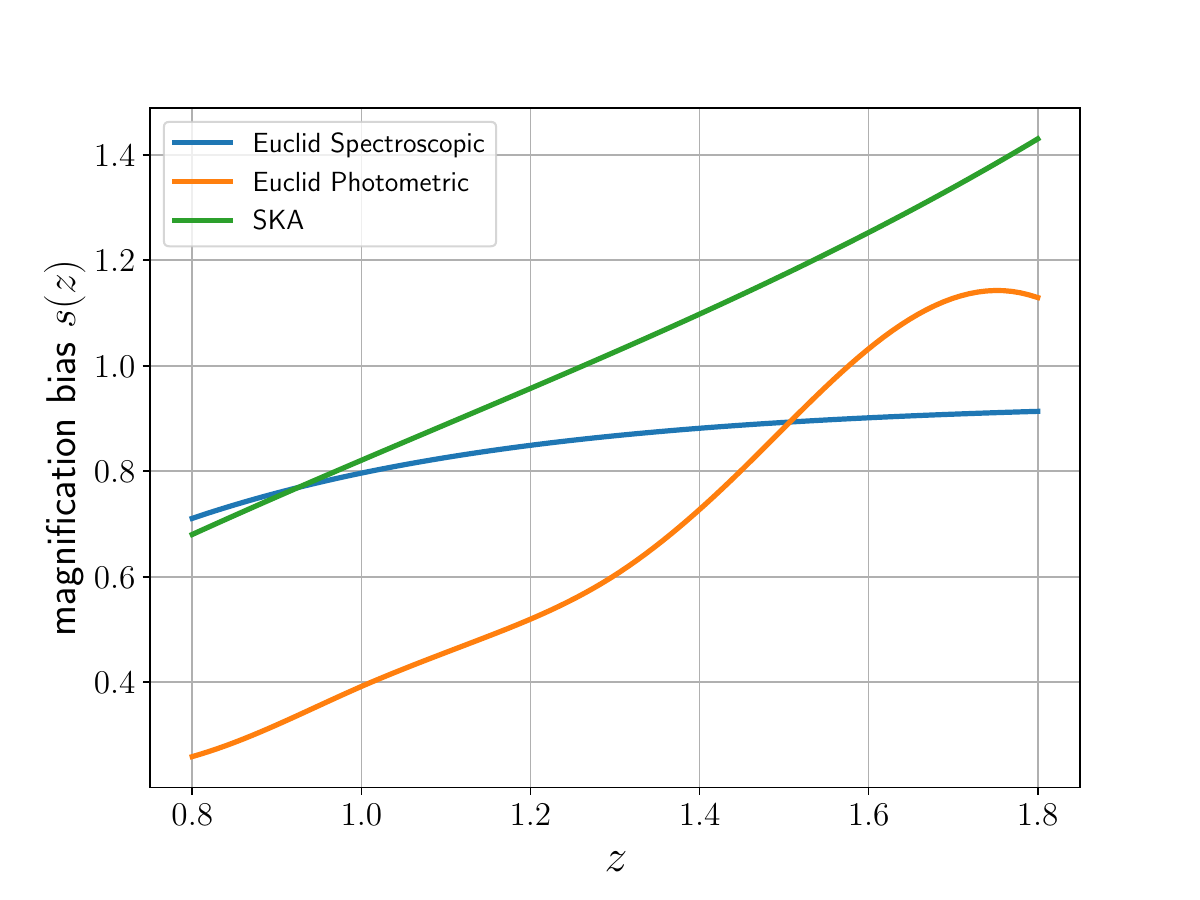}
    \caption{\emph{Left panel}: Galaxy bias $b_\g(z)$ for \emph{Euclid}-like spectroscopic, \emph{Euclid}-like photometeric and SKA2-like surveys. We also show intensity mapping bias for HIRAX. \emph{Right panel: }magnification bias for \emph{Euclid}-like spectroscopic, \emph{Euclid}-like photometeric and SKA2-like.}
    \label{fig:gal-mag-bias}
\end{figure*}

Galaxy and magnification biases are plotted in Fig.~\ref{fig:gal-mag-bias}.

\section{Window Functions}
We use the top-hat window function for spectroscopic surveys and a Gaussian window for the photometric survey. The reason is that in spectroscopic surveys, we can measure precisely the redshift, while in photometric surveys, we have larger errors on the redshift. For the top-hat, we choose a smooth version of the top-hat filter:
\begin{equation}
W_{\text{TH}}(z) = \begin{cases}
\frac{1}{2}\Big[1+\tanh\Big(\frac{z - \mu+\Delta z /2}{\alpha}\Big)\Big] & \text{if } z \leq \mu\\
\frac{1}{2}\Big[1+\tanh\Big(-\frac{(z - \mu-\Delta z /2)}{\alpha}\Big)\Big] & \text{otherwise}
\end{cases}
\end{equation}
where $\mu$ is the mean of the redshift bin, $\Delta z$ is the width of the bin, and $\alpha$ is the speed of the transition which we choose to be $\alpha=\Delta z/10$. The expression for the Gaussian window function is given by
\begin{equation}
W_{\text{G}}(z) = \frac{1}{\Delta z\sqrt{2\pi}}\exp\left[-\frac{(z-\mu)^2}{2\Delta z^2}\right]\, .
\end{equation}

\bibliographystyle{unsrt}
\bibliography{gimco_Egbib}

\end{document}